\documentclass[plb,preprintnumbers,nofootinbib]{elsarticle}
\usepackage{bm}
\usepackage{latexsym}
\usepackage{dcolumn}
\usepackage{amsmath,amsfonts,amssymb}
\usepackage{graphicx,epsfig}
\usepackage{color}
\usepackage{amsthm}

\usepackage{lineno}

\usepackage[citecolor=purple,linkcolor=blue,urlcolor=blue,colorlinks=true,filecolor=green]{hyperref}

\usepackage[utf8x]{inputenc}
\usepackage{mathtools}
\usepackage{amsfonts}
\usepackage{amsmath}
\usepackage{amssymb,epsf}
\usepackage{latexsym}
\usepackage{graphicx,epsfig}
\usepackage{amssymb}
\usepackage{subfigure}

\newcommand*{\Scale}[2][4]{\scalebox{#1}{$#2$}}%

\usepackage{epstopdf}
\usepackage{color}
\usepackage[normalem]{ulem}
\usepackage{slashed}

\usepackage{pdfpages}

\usepackage{eqnarray,amsmath,amssymb,mathrsfs,multirow,color}
\usepackage{amsmath}
\usepackage{amsfonts}

\usepackage{makecell}

\usepackage{graphicx}
\usepackage{graphicx}
\graphicspath{ {images/} }
\usepackage{commath}
\usepackage{amsthm,amssymb,amsmath,mathrsfs}
\usepackage{mathtools}
\usepackage{empheq}
\usepackage{tabu} 
\usepackage[most]{tcolorbox}
\usepackage{fixmath}

\DeclarePairedDelimiterX\braket[2]{\langle}{\rangle}{#1 \delimsize\vert #2}
\graphicspath{ {images/} }

\def\be {\begin{equation}}
	\def\ee {\end{equation}}
\def\bea {\begin{eqnarray}}
	\def\eea {\end{eqnarray}}
\def\bc {\begin{center}}
	\def\ec {\end{center}}
\def\bg {\begin{align}}
	\def\eg {\end{align}}
\def\bi {\begin{itemize}}
	\def\ei {\end{itemize}}
\def\bm {\begin{pmatrix}}
	\def\em {\end{pmatrix}}

\def\le {\left}
\def\ri {\right}
\def\p {\partial}

\def\l {\ell}

\def\1{_{_1}}
\def\2{_{_2}}

\def\c  {\cdot}

\def\g  {\gamma}
\def\d  {\delta}

\def\e  {\eta}

\def\m  {\mu}
\def\n  {\nu}

\def\w {\omega}

\begin{document}                             
	
	\begin{frontmatter}
		
		\title{{\bf Massive to Massless \\\vspace{.2cm} by \\\vspace{.2cm} Applying a Nonlocal Field Redefinition}}
		

		\author[IPM,Mashhad]{Mojtaba \textsc{Najafizadeh} \vspace{.2cm}}
		
		
		\address[IPM]{\,School of Physics, Institute for Research in Fundamental Sciences (IPM),\\ P.O.Box 19395-5531, Tehran, Iran\vspace{.3cm}}

		\address[Mashhad]{\,Department of Physics, Faculty of Science, Ferdowsi University of Mashhad,\\ P.O.Box 1436, Mashhad, Iran\\\vspace{.3cm}
		{\color{blue}{\normalsize mnajafizadeh@ipm.ir}}}
		
		
\begin{abstract}
	The invariance of physical observables like particle's mass under a {\bf local} field redefinition is a well-known and important property of quantum field theory. In this paper, on the other hand, we investigate {\bf nonlocal} field redefinitions eliminating a physical observable from a theory. Indeed, by giving explicit examples, we present invertible nonlocal field redefinitions reducing a massive theory to the massless one, and vice versa. The studied examples are: classical harmonic oscillator, the scalar field theory, and the Dirac field theory.


	
\end{abstract}
		
		\begin{keyword}
			{\small Nonlocal field redefinition, Harmonic oscillator, Scalar field, Dirac field
			\\\vspace{.1cm}
			\texttt{Preprint code:IPM/P-2023/07}}
		\end{keyword}
		
	\end{frontmatter}
	
	
	\newpage
	
	\section{Introduction} \label{1}
	Some of coupling parameters in any action principle can be removed by a redefinition of the fields, which are known as the redundant parameters \cite{Weinberg:1995mt}. On the other hand, physical observables like particle masses and scattering amplitudes are independent of field redefinitions which is a well-known and important property of quantum field theory. In order to clarify the issue, let us follow Weinberg's approach \cite{Weinberg:1995mt} and consider the action of a scalar field theory in the form
	\be 
	S=\int d^4 x\le[-\tfrac{\,1\,}{2}\,Z\,(\p^\m\mathrm{\Phi}\,\p_\m\mathrm{\Phi}+m^2\mathrm{\Phi}^2)-\tfrac{1}{4!}\,g\,Z^2\mathrm{\Phi}^4\ri].\label{action g}
	\ee 
	The constant $Z$ which is called a field renormalization constant is a redundant coupling because $\d S\Scale[0.9]{/}\d Z$ vanishes when we use the field equation. In other words, the constant $Z$ can be removed by redefining the field, $\mathrm{\Phi}\rightarrow (\Scale[.85]{{1}/{\sqrt{Z}}})\,\mathrm{\Phi}$, while neither the scalar particle mass $m$ nor the coupling $g$ can be eliminated by a field redefinition. Indeed, in this example, the field redefinition is continuous and local, but what happens if one applies a nonlocal field redefinition? 
	
	\vspace{.3cm}
	
	In this paper, we observe by applying nonlocal field redefinition a physical parameter of a theory may be redundant as well. The physical parameter we aim to make redundant is angular frequency of the simple harmonic oscillator, as well as mass of the scalar field and the Dirac field. As we know, in the last two examples, massive fields of spin-$0$ and spin-$\tfrac{1}{2}$ have the same degrees of freedom as their corresponding massless fields, that is, in four dimensions, 1 for the massive$\Scale[0.8]{/}$massless real scalar field and 8 for the massive$\Scale[0.8]{/}$massless Dirac field. However, except these two fields, other arbitrary fields of spin $s \,(\,\geqslant1)$ have different degrees of freedom depending on being massive or massless\footnote{\,See e.g. \cite{Rahman:2012thy} for degrees of freedom of a massive$\Scale[0.8]{/}$massless arbitrary spin-$s$ field in any spacetime dimension.}. Let us refer to these cases in the conclusion.

	
	\vspace{.3cm}
	
	Therefore, the present work is devoted to investigating a set of field redefinitions subject to nonlocality. The layout of the paper is as follows. In Sec. \ref{2}, we study the classical harmonic oscillator which simplifies the calculations in the next two sections. In Secs. \ref{3} and \ref{4}, we investigate the nonlocality of field redefinitions for the scalar field theory and the Dirac field theory respectively. We conclude and present open problems in Sec. \ref{5}. 
	
	\vspace{.3cm}
	
	We will work in the``mostly plus'' signature for the metric and concentrate on $d$-dimensional Minkowski spacetime. Natural units, $c=1=\hbar$, and shorthand notation $\p_\m:=\p\Scale[0.85]{/}\p x^\m$ (with $\m=0,1,\ldots,d-1$) will be used. 
	
	\section{Harmonic Oscillator} \label{2}
	Although our concentration may lie on the classical$\Scale[0.8]{/}$quantum field theories, it is beneficial to first study classical harmonic oscillator shedding light on our nonlocal field redefinitions discussed in Secs. \ref{3} and \ref{4}. Thus, let us consider a Newtonian massive particle whose position only depends on time and subject to the force law of the usual Hooke's law. The equation of motion reads
	\be
	\Big(\, \frac{d^2}{dt^2}\,+\,\w^2\,\Big)\,\mathrm{X}(t)=0\,,\label{Osci1}
	\ee
	where $\w=\Scale[0.9]{\sqrt{{k}/{m}}}\,,\, m\neq0,$ is angular frequency of a simple harmonic oscillator system. In comparison to the given example in Sec. \ref{1}, here, the angular frequency is not a redundant parameter as it is a physical observable and cannot be removed by a so-called local wave function redefinition. Nevertheless, it may be redundant by performing a nonlocal wave function redefinition. To see this redundancy, we introduce a new wave function $x(t)$ through
	\begin{tcolorbox}[ams align, colback=white!95!purple]
	\mathrm{X}(t)=\bigg(\,\,\underset{n=0}{\overset{\infty}{\textstyle\sum}}\,(\,\w t\,)^{2n}\,~\frac{\Scale[0.85]{(-1)}^n}{4^n~n!~(t\,\tfrac{d}{dt}+\tfrac{1}{2})_n}\,\,\bigg)\,x(t)\,, \label{field red1}
	\end{tcolorbox}
	\noindent where $(\cdots)_n$ in the denominator stands for the rising Pochhammer symbol \cite{Abramowitz} with the argument $a$ of any $n \in \mathbb{N}$ 
	\be
	(a)_n=a(a+1)\cdots(a+n-1)\,. \label{pochhammer}
	\ee 
	By plugging \eqref{field red1} into the equation of motion \eqref{Osci1}, and doing some calculations explained below, we shall arrive at 
	\be 
	\Big(\,\frac{d^2}{dt^2}\,\Big)\,x(t)=0\,.\label{free1}
	\ee
	Multiplying the latter by $m$, $m\neq0$, it is nothing but a free Newtonian massive particle subject to no force. As a consequence, the nonlocal wave function redefinition \eqref{field red1} could drop the physical observable, that is angular frequency, from the original equation \eqref{Osci1}. It is remarkable to mention that relation \eqref{field red1} can be also expressed in terms of the Bessel function of the first kind (see e.g. \cite{Najafizadeh:2017tin}). However, we shall keep the above form dealing with the Pochhammer symbol properties which makes for easy calculations, as we will see later.
	
	\vspace{.3cm}
	
	We have found that the above procedure can be invertible. In fact, this time, we consider the free Newtonian massive particle equation \eqref{free1} as our starting point. After that, we introduce a new wave function $\mathrm{X}(t)$ via 
	\begin{tcolorbox}[ams align, colback=white!95!purple]
	x(t)=\bigg(\,\,\underset{n=0}{\overset{\infty}{\textstyle\sum}}~\frac{\Scale[0.85]{1}}{~4^n\,n!\,(t\,\tfrac{d}{dt}-n-\tfrac{1}{2})_n}~(\,\w t\,)^{2n}\,\bigg)\,\mathrm{X}(t)\,.\label{redef invert}
	\end{tcolorbox} 
	\noindent By plugging the latter into \eqref{free1} and performing some calculations we can reach the simple harmonic oscillator equation \eqref{Osci1}. This means as the nonlocal wave function redefinition could remove the angular frequency from the system, invertibly, we could restore it to the system by an inverse operation. Therefore, the angular frequency as a physical observable could be redundant upon applying a nonlocal wave function redefinition.

	\vspace{.3cm}
	
	Let us now go into the details of the above calculations. First of all, we note that the appeared operators in \eqref{Osci1}-\eqref{redef invert}, which are ${d^2}/{dt^2}$, $t^2$ and $t\,{d}/{dt}$, can be packed into the following generators
	\be 
	\l_0:=\tfrac{1}{2}\,(t\,\tfrac{d}{dt}+\tfrac{1}{2})\,,\quad\quad\quad
	\l_1:=-\,\tfrac{1}{2}\,\tfrac{d^2}{dt^2}\,, \quad\quad\quad
	\l_{-1}:=-\,\tfrac{1}{2}\,t^2\,.\nonumber
	\ee  
	These generators satisfy the $sl(2,R)$ algebra
	\be 
	[\,\l_m\,,\,\l_n\,]=(m-n)\,\l_{m+n}\,, \qquad m,n=0,\pm 1\,.
	\ee 
	It is then convenient to find the following commutation relation for any $n \in \mathbb{N}$  
	\be 
	\Big[\,\,\frac{d^2}{dt^2}\,,\,t^{2n}\,\,\Big]=4n\,(a-n+1)\,t^{2n-2}\,, \label{comm}
	\ee
	where 
	\be
	a:=t\,\frac{d}{dt}\, + \frac{1}{2}\,.\label{a}
	\ee
	At this moment, let us define two nonlocal dimensionless (\,$[\,t\,]=\mathrm{T}=[\w]^{-1}$) operators 
	\begin{align}
		\mathrm{P}&:= \underset{n=0}{\overset{\infty}{\textstyle\sum}}\,(\,\w t\,)^{2n}\,~\frac{\Scale[0.85]{(-1)}^n}{4^n~n!~(a)_n}\,,\label{P}\\[5pt]
		\mathrm{Q}&:= \underset{n=0}{\overset{\infty}{\textstyle\sum}}\,(\,\w t\,)^{2n}\,~\frac{\Scale[0.85]{(-1)}^n}{4^n~n!~(a+2)_n}\,,
	\end{align} 
	where $(a)_n$ and $a$ were respectively introduced in \eqref{pochhammer}, \eqref{a}. We recall that the former operator \eqref{P} appeared in \eqref{field red1}. At this stage, using the commutation relation \eqref{comm}, and the property of the Pochhammer symbol \cite{Abramowitz}
	\be 
	(a)_n=(a+n-1)\,(a)_{n-1}\,, \label{property}
	\ee 
	it is straightforward to examine that two operators, $\mathrm{P}$ and $\mathrm{Q}$, will satisfy the following identity
	\be 
	\Big(\,\frac{d^2}{dt^2}\,+\,\w^2\,\Big)\,\mathrm{P}=\mathrm{Q}\,\Big(\,\frac{d^2}{dt^2}\,\Big)\,.\label{Identity}
	\ee
	As a result, by applying the nonlocal wave function redefinition \eqref{field red1}, or equivalently $\mathrm{X}=\mathrm{P}\, x$, and then using the identity \eqref{Identity}, we immediately realize that Eq. \eqref{Osci1} can be reduced to \eqref{free1}, i.e.
	\be 
	\Big(\,\frac{d^2}{dt^2}\,+\,\w^2\,\Big)\mathrm{X}=0
	~~~~\rightarrow~~~~
	\Big(\,\frac{d^2}{dt^2}\,+\,\w^2\,\Big)\mathrm{P}\,x=0
	~~~~\rightarrow~~~~
	\mathrm{Q}\,\Big(\,\frac{d^2}{dt^2}\,\Big)x=0\,.\nonumber
	\ee

	
	

    Let us end this section by discovering inverse operators. This enables us to follow the reverse procedure applied above. One way to find the inverse operators is as follows. Using the fact that $\mathrm{P}\mathrm{P}^{-1}=1=\mathrm{Q}^{-1}\mathrm{Q}$, we can multiply the identity \eqref{Identity} by $\mathrm{Q}^{-1}$ to the left and by $\mathrm{P}^{-1}$ to the right which results in 
	\be 
	\mathrm{Q}^{-1}\Big(\,\frac{d^2}{dt^2}\,+\,\w^2\,\Big)=\Big(\,\frac{d^2}{dt^2}\,\Big)\mathrm{P}^{-1}\,.\label{Identity inverse}
	\ee  
	Here we ask ourselves if it is possible to figure out inverse operators, $\mathrm{P}^{-1}$ and $\mathrm{Q}^{-1}$, such that satisfy \eqref{Identity inverse}? The answer is positive. In fact, if we introduce
	\begin{align}
		\mathrm{P}^{-1}&:= \underset{n=0}{\overset{\infty}{\textstyle\sum}}~\frac{\Scale[0.85]{1}}{~4^n~n!~(a-n-1)_n}~(\,\w t\,)^{2n}\,,\label{P-1}\\[5pt]
		\mathrm{Q}^{-1}&:= \underset{n=0}{\overset{\infty}{\textstyle\sum}}~\frac{\Scale[0.85]{1}}{~4^n~n!~(a-n+1)_n}~(\,\w t\,)^{2n}\,,
	\end{align}
    the so-called ``inverse identity'' \eqref{Identity inverse} will be then satisfied. To check this, it is sufficient to use \eqref{comm} and \eqref{property}. We note that the inverse operator in \eqref{P-1} is the one that became visible in \eqref{redef invert}. Therefore, in a reverse way, the Eq. \eqref{free1} could be reduced to \eqref{Osci1}, upon taking into account the nonlocal wave function redefinition \eqref{redef invert}, corresponding to $x= \mathrm{P}^{-1}~\mathrm{X}$, and applying the inverse identity \eqref{Identity inverse}. Inspired by the harmonic oscillator, let us begin studying the classical$\Scale[0.8]{/}$quantum field theory in the next two sections.

	\section{Scalar field theory} \label{3}
	We take into account a real scalar field theory in $d$-dimensional Minkowski spacetime given by the action 
	\be 
	S=\tfrac{1}{2}\int d^dx\,\,\mathrm{\Phi}\,\big(\,\p^\m\p_\m-m^2\,\big)\,\mathrm{\Phi}\,,\label{scalar action}
	\ee 
	where $m$ is scalar field's mass and $\m=0,1,\ldots,d-1$. We note, to simplify calculations, an integration by part is used by which the standard field derivative term, $-\,\p^\m \mathrm{\Phi} \,\p_\m\mathrm{\Phi}$ ($\,=\mathrm{\Phi}\,\p^\m\p_\m\mathrm{\Phi}\,+\,\hbox{total derivative}$), has been rewritten as the above form. As already mentioned, the mass parameter in the action \eqref{scalar action} is not redundant and does not eliminate by a local field redefinition \cite{Weinberg:1995mt}. Nevertheless, we can make it redundant if we introduce a nonlocal field redefinition, which is
	\begin{tcolorbox}[ams align, colback=white!95!purple] 
	\mathrm{\Phi}=\bigg(\,\underset{n=0}{\overset{\infty}{\textstyle\sum}}\,(\,m x\,)^{2n}\,~\frac{\Scale[0.85]{(-1)}^n}{4^n\,n!~\big(x^\m\,\p_\m+\tfrac{d}{2}\big)_n}\,\bigg)\,\varphi\,. \label{scalar field red}
	\end{tcolorbox}
\noindent Indeed, by plugging the latter into \eqref{scalar action}, the massive scalar field action \eqref{scalar action} reduces to (details of the calculations are given below)
	\be 
	\mathcal{S}=\tfrac{1}{2}\int d^dx\,\,\varphi\,\p^\m\p_\m\,\varphi\,,\label{scalar action free}
	\ee
	which is a massless scalar field action. We notice the action \eqref{scalar action free} is conformal invariant and the scaling dimension of $d$-dimensional scalar field is $\mathrm{\Delta}=(d-2)/2$.
	
		\vspace{.3cm}
	
	Invertibly, we discover the massless scalar field action \eqref{scalar action free} can be reduced to the massive one \eqref{scalar action} by applying the following nonlocal field redefinition  
	\begin{tcolorbox}[ams align, colback=white!95!purple]  
	\varphi=\bigg(\,\underset{n=0}{\overset{\infty}{\textstyle\sum}}~\frac{\Scale[0.85]{1}}{~4^n\,n!~\big(x^\m\p_\m-n+\tfrac{d-2}{2}\big)_n}~(m x)^{2n}\,\bigg)\,\mathrm{\Phi}\,. \label{scalar field red2}
	\end{tcolorbox}
	\noindent Here, we note that the operator $x^\m\p_\m$ in the denominator, denoting by $\mathrm{D}$, is nothing but the dilatation (rigid scaling) operator. Hence, the argument of the Pochhammer symbol in \eqref{scalar field red2} can be taken into account as $\mathrm{D}-n+\mathrm{\Delta}$. As we will see later, we expect a similar structure for the Dirac field as well.
	
	\vspace{.3cm}
	
	In this place, we give useful relations enabling the reader to simply check the details of calculations. To this end, we first obtain the following commutation relation, for any $n \in \mathbb{N}$, in any spacetime dimension $d$
	\be 
	\le[\,\,\p^\m\p_\m\,,\,x^{2n}\,\,\ri]=4n\,(h-n+1)\,x^{2n-2}\,, \label{comm 1}
	\ee
	where 
	\be
	h:=x^\m\,\p_\m+\frac{d}{2}\,.\label{h}
	\ee 
	After that, we define dimensionless (\,$[m]=\mathrm{M}=[x]^{-1}$) operators in $d$ spacetime dimensions 
		\begin{align}
		&\mathbf{P}:=\underset{n=0}{\overset{\infty}{\textstyle\sum}}\,(\,m x\,)^{2n}\,~\frac{\Scale[0.85]{(-1)}^n}{4^n\,n!~(h)_n}\,,\label{p}\\[8pt]
		&\mathbf{Q}:=\underset{n=0}{\overset{\infty}{\textstyle\sum}}\,(\,m x\,)^{2n}\,~\frac{\Scale[0.85]{(-1)}^n}{4^n\,n!~(h+2)_n}\,,\label{q}\\[8pt]
		&\mathbf{P}^{-1}:=\underset{n=0}{\overset{\infty}{\textstyle\sum}}\,~\frac{\Scale[0.85]{1}}{4^n\,n!~(h-n-1)_n}~(\,m x\,)^{2n}\,,\\[8pt] &\mathbf{Q}^{-1}:=\underset{n=0}{\overset{\infty}{\textstyle\sum}}\,~\frac{\Scale[0.85]{1}}{4^n\,n!~(h-n+1)_n}~(\,m x\,)^{2n}\,,\label{q-1}
	\end{align}
	where the Pochhammer symbol $(h)_n$ and $h$ were introduced in \eqref{pochhammer}, \eqref{h}. We note that the operators $\mathbf{P}$ and $\mathbf{P}^{-1}$ are those employed in \eqref{scalar field red} ($\mathrm{\Phi}=\mathbf{P}\,\varphi$) and \eqref{scalar field red2} ($\varphi=\mathbf{P}^{-1}\,\mathrm{\Phi}$) respectively. Now, using the commutation relation \eqref{comm 1} and the Pochhammer symbol property \eqref{property}, we can conveniently demonstrate that the set of operators \eqref{p}-\eqref{q-1} will satisfy the following identities
	\begin{align} 
	\big(\,\p^\m\p_\m-m^2\,\big)\,\mathbf{P}&=\mathbf{Q}\,\big(\,\p^\m\p_\m\,\big)\,,\label{11}\\[8pt]
	\mathbf{Q}^{-1}\,\big(\,\p^\m\p_\m-m^2\,\big)&=\big(\,\p^\m\p_\m\,\big)\,\mathbf{P}^{-1}\,.\label{22}
    \end{align}  
    Furthermore, since we argue at the level of the action, Hermitian conjugation of the above operators is needed to calculate. Therefore, we introduce the Hermitian conjugation rules  
    \be 
    (x^\m)^\dagger=x^\m\,,\qquad (\p_\m)^\dagger=-\,\p_\m\,,\label{herm}
    \ee 
    under which, e.g., $h^\dagger=-h$. Using \eqref{herm}, we can perform some calculations and realize that the Hermitian conjugation of \eqref{p} and \eqref{q} results in 
    \be
    \mathbf{P}^\dagger=\mathbf{Q}^{-1}\,, \qquad \mathbf{Q}^\dagger=\mathbf{P}^{-1}\,. \label{p dage}
    \ee  
    These results could be also obtained without any explicit calculations from \eqref{11}, \eqref{22}. In fact, by taking the Hermitian conjugation of \eqref{11}, and its comparison with \eqref{22}, the relations in \eqref{p dage} will be acquired directly. 
    
    \vspace{.3cm}
    
    Using the above facilities, one can conveniently reduce the massive action to the massless one and vice versa. For instance, applying the field redefinition \eqref{scalar field red}, which is corresponding to $\mathrm{\Phi}=\mathbf{P}\,\varphi=\varphi\,\mathbf{P}^\dagger$, the action \eqref{scalar action} reduces to \eqref{scalar action free}
    \begin{align}
    	S=\tfrac{1}{2}\int d^dx\,\,\varphi\,\mathbf{P}^\dagger\big(\,\p^\m\p_\m-m^2\,\big)\,\mathbf{P}\,\varphi=\tfrac{1}{2}\int d^dx\,\,\varphi\!\underbrace{\,\mathbf{P}^\dagger\,\mathbf{Q}\,}_{=\,1}\big(\,\p^\m\p_\m\,\big)\,\varphi=\mathcal{S}\,,\nonumber
    \end{align}
    in which relations \eqref{11}, \eqref{p dage} have been used.

	
	\section{Dirac field theory} \label{4}
	In $d$-dimensional flat spacetime, the massive Dirac field is given by the action
	\be
	S=-\int d^dx\,\bar{\mathrm{\Psi}}\big(\,\gamma\c\p+m\,\big)\mathrm{\Psi}
	\label{massive dirac}
	\ee 
	where the dot stands for the scalar product, $\g^\m$ are $d$-dimensional Dirac gamma matrices, and $\bar{\mathrm{\Psi}}:={\mathrm{\Psi}}^\dagger\,i\,\g^0$ is the Dirac adjoint. Similar to the scalar field, here, we intend to illustrate that the physical mass parameter $m$ can be redundant, by applying a nonlocal field redefinition. For this purpose, we introduce a new field $\psi$ by 
	\begin{tcolorbox}[ams align, colback=white!95!purple]  
\hspace{-.5cm}	\mathrm{\Psi}=\bigg(\,\underset{k=0}{\overset{\infty}{\textstyle\sum}}\,\Big[(m\,\gamma \cdot x)^{2k} - 2k\, (m\,\gamma \cdot x)^{2k-1}\Big]
		\,\frac{\Scale[0.85]{1}}{~4^k~k!~\big(x^\m\,\p_\m+\tfrac{d}{2}\big)_k~}\bigg)\,\psi\,.\label{Psii}
	\end{tcolorbox}
\noindent Plugging the latter into the massive action \eqref{massive dirac}, and doing some calculations explained below, the mass parameter will be eliminated so as we shall arrive at 
	\be 
	\mathcal{S}=-\int d^dx~\bar{\psi}~\gamma\c\p~{\psi}\,,
	\label{massless dirac}
	\ee 
that is the Dirac action for the massless field $\psi$.


As a matter of fact, we realized that this method can be invertible as well. Indeed, if we introduce the nonlocal field redefinition
	\begin{tcolorbox}[ams align, colback=white!95!purple]  
\hspace{-.6cm}	\psi=\bigg(\,\underset{k=0}{\overset{\infty}{\textstyle\sum}}\,\frac{\Scale[0.85]{(-1)}^k}{4^k\,k!\,\big(x^\m\,\p_\m-k+\tfrac{d}{2}\big)_k~}\Big[(m\,\gamma \cdot x)^{2k} - 2k\, (m\,\gamma \cdot x)^{2k-1}\Big]\bigg)\,\mathrm{\Psi},\label{psii}
	\end{tcolorbox}
\noindent then the mass parameter can be restored to the theory, such that this time by plugging \eqref{psii} into the massless action \eqref{massless dirac}, we can arrive back to the massive action \eqref{massive dirac}. We note that, similar to the scalar field redefinition \eqref{scalar field red2}, here, the argument of the Pochhammer symbol in \eqref{psii} can be considered as $\mathrm{D}-k+{\varDelta}+\tfrac{1}{2}$, where $\varDelta=(d-1)/2$ is the scaling dimension of $d$-dimensional Dirac spinor field.   
	
	\vspace{.3cm}
	
	The step-by-step details of the calculations are very similar to the case of the scalar field. However, here, we have the Clifford algebra $\{\g^\m,\g^\n\}=2\e^{\m\n}$ that using it one can simply find $\{\g\c\p,\g\c x\}=2(x\c\p)+d$ and $[\g\c\p,(\g\c x)^2]=2(\g\c x)$. In the same way, one can find a generalization of odd and even powers of $\g\c x$. Therefore, we acquire commutation ($[a, b] = ab-ba$) and anticommutation ($\{a, b\} = ab + ba$) relations, in any spacetime dimension $d$, for any $k \in \mathbb{N}$
	\bea
	\le[\,\gamma \cdot \p ~, ~(\gamma \cdot x)^{\,2k} \,\ri] &=&2k\, (\gamma \cdot x)^{\,2k-1} \,,\label{comu}  \\[5pt]
	\le\{\gamma \cdot \p \,,\, (\gamma \cdot x)^{\,2k-1} \ri\} &=&2\,(h-k+1)~ (\gamma \cdot x)^{\,2k-2} \,,\label{anti}
	\eea
	where $h$ defined in \eqref{h}. After that, we introduce dimensionless operators in $d$ spacetime dimensions
	\begin{align}
		\mathbb{P} &:=\underset{k=0}{\overset{\infty}{\textstyle\sum}}\,\Big[(m\,\gamma \cdot x)^{2k} - 2k\, (m\,\gamma \cdot x)^{2k-1}\Big]
		\frac{\Scale[0.85]{1}}{~4^k~k!~(h)_k~}  \,,\label{p bb}\\[5pt]
		\mathbb{Q} &:=\underset{k=0}{\overset{\infty}{\textstyle\sum}}\,\Big[(m\,\gamma \cdot x)^{2k} + 2k\, (m\,\gamma \cdot x)^{2k-1}\Big]
		\frac{\Scale[0.85]{1}}{~4^k~k!~(h+1)_k~}\,, \label{q bb}\\[5pt]
		\mathbb{P}^{-1}&:=\underset{k=0}{\overset{\infty}{\textstyle\sum}}\,\frac{\Scale[0.85]{(-1)}^k}{~4^k~k!~(h-k)_k~}\,\Big[(m\,\gamma \cdot x)^{2k} - 2k\, (m\,\gamma \cdot x)^{2k-1}\Big]
		 \,,\\[5pt]
		\mathbb{Q}^{-1}&:=\underset{k=0}{\overset{\infty}{\textstyle\sum}}\,\frac{\Scale[0.85]{(-1)}^k}{~4^k~k!~(h-k+1)_k~}\,\Big[(m\,\gamma \cdot x)^{2k} + 2k\, (m\,\gamma \cdot x)^{2k-1}\Big]\,,\label{Q-1}
	\end{align}
	with $h$ introduced in \eqref{h}. We notice that the operators $\mathbb{P}$ and $\mathbb{P}^{-1}$ are those emerged in \eqref{Psii} ($\mathrm{\Psi}=\mathbb{P}\,\psi$) and \eqref{psii} ($\psi=\mathbb{P}^{-1}\,\mathrm{\Psi}$). Using the commutation \eqref{comu} and anti-commutation \eqref{anti} relations as well as the property of the Pochhammer symbol \eqref{property}, it follows that the above operators satisfy the following identities
	\begin{align} 
		\big(\,\gamma\c\p+m\,\big)\,\mathbb{P}&=\mathbb{Q}\,\big(\,\gamma\c\p\,\big)\,,\label{111}\\[8pt]
		\mathbb{Q}^{-1}\,\big(\,\gamma\c\p+m\,\big)&=\big(\,\gamma\c\p\,\big)\,\mathbb{P}^{-1}\,.\label{222}
	\end{align} 
	To find the Hermitian conjugate of operators, we use \eqref{herm} accompanied with
	\be 
	(\g^\m)^\dagger=\g^0\g^\m\g^0\,,\qquad (\g^0)^\dagger=-\,\g^0\label{hermit}\,, \qquad (\g^0)^2=-1\,.
	\ee 
    As a consequence, the Hermitian conjugates of the operators \eqref{p bb}-\eqref{Q-1} read 
	\begin{align}
	\mathbb{P}^\dagger&=-\,\g^0\,\mathbb{Q}^{-1}\,\g^0\,, \qquad\qquad
	(\,\mathbb{P}^{-1})^\dagger=-\,\g^0\,\mathbb{Q}\,\g^0\,, \label{pb dage} \\[5pt]
	 \mathbb{Q}^\dagger&=-\,\g^0\,\mathbb{P}^{-1}\,\g^0\,, \qquad\qquad
	(\mathbb{Q}^{-1})^\dagger=-\,\g^0\,\mathbb{P}\,\g^0\,. 
	\end{align} 
	
	
	At this stage, we are equipped to reduce each actions, \eqref{massive dirac} or \eqref{massless dirac}, to each other by applying an appropriate nonlocal field redefinition. For instance, let us apply $\psi=\mathbb{P}^{-1}\,\mathrm{\Psi}$ \eqref{psii} (which, using \eqref{pb dage}, its Dirac adjoint becomes $\bar\psi=\bar{\mathrm{\Psi}}\,\mathbb{Q}$) into the massless Dirac field action \eqref{massless dirac} arriving at \eqref{massive dirac}. It reads
	\begin{align}
		\mathcal{S}&=-\int d^dx~\bar{\mathrm{\Psi}}\,\mathbb{Q}~(\gamma\c\p)~\mathbb{P}^{-1}\,\mathrm{\Psi}
		=-\int d^dx~\bar{\mathrm{\Psi}}\,\underbrace{\mathbb{Q}\,\mathbb{Q}^{-1}}_{=\,1}(\gamma\c\p+m)~\mathrm{\Psi}=S\,,\nonumber
	\end{align}
	in which relations \eqref{222} and \eqref{pb dage} have been utilized.

	\section{Conclusions and Discussion} \label{5}
	
	In the quantum field theory, redundant coupling parameters can be removed by applying a local field redefinition, while physical observables like particle masses and scattering amplitudes are independent of field redefinitions \cite{Weinberg:1995mt}. In this paper, we investigated the possibility if a physical mass parameter can be redundant by applying a nonlocal field redefinition. For this purpose, we presented three explicit examples equipped with invertible nonlocal wave$\Scale[0.8]{/}$field redefinitions. In the case of the classical harmonic oscillator, the angular frequency became redundant  using \eqref{field red1}, while it could be recaptured to the system by \eqref{redef invert}. For the scalar field theory, by applying \eqref{scalar field red}, the mass parameter of the spin-$0$ field was removed from the theory, while it could be reattached to the theory upon \eqref{scalar field red2}. In similar fashion, we introduced \eqref{Psii} and its inverse \eqref{psii} for the Dirac field theory, using them, the mass parameter of the spin-$\tfrac{1}{2}$ spinor field could be eliminated and restored respectively.
	
	\vspace{.3cm} 
	
	The present work opens several avenues for further research. For instance, as already mentioned in the introduction, except spin-$0$ and spin-$\tfrac{1}{2}$ fields, other arbitrary fields of spin $s \,(\,\geqslant1)$ have different degrees of freedom depending on being massive or massless. For example, in $d$ dimensions, the spin-$1$ field has ($d-1$) massive degrees of freedom, while it has ($d-2$) massless degrees of freedom. Thus, it is of interest to find invertible operators (if any) making the mass parameter redundant in the Proca action in which the degrees of freedom will change by mass redundancy. If such operators exist, another example for future investigations is studying mass redundancy in the massive Fierz-Pauli action \cite{Fierz:1939ix} as well as massive higher spin field theories\footnote{\,In this case, if one finds inverse operators, then, the argument of the Pochhammer symbol may be expected to be of the form $\mathrm{D}-n+{\mathrm{\Delta}}+s$, where $s$ is spin of the field.} (see e.g. \cite{Buchbinder:2005ua, Metsaev:2006zy}), in which it seems that something like the Stueckelberg mechanism should play a role\footnote{\,We thank Sylvester James Gates Jr. for bringing our attention to this point.}.

	\vspace{.3cm}
	
	Moreover, in this work, we concentrated on the mass redundancy in a free theory. Further investigation can eliminate the mass parameter in an interacting theory, for example in \eqref{action g}, for which the limit $g\rightarrow0$ should recover our results. An even deeper study would remove the coupling parameter in an interacting theory. For this case, one may refer to \cite{Nicolai:1979nr,Nicolai:1980jc} including nonlocal (and nonlinear) transformations that map an interacting theory to a free one, and where the fermions are needed to match the Jacobian of the transformation (see also \cite{Malcha:2021ess} and references therein for a recent development). 
	
	\vspace{.3cm}
	
		Furthermore, our method may establish a duality between theories in flat and curved spacetimes. To be more precise, the scalar field theory in anti-de Sitter spacetime includes the cosmological constant playing a role like mass. Thus, making use of our nonlocal transformations, one can simply drop the cosmological constant from the field equation which this in turn yields the equation of motion in flat spacetime. Therefore, as an application, a suitable generalization of our nonlocal transformations may be beneficial to construct a flat spacetime holographic correspondence \cite{Pasterski:2021raf}.

	\vspace{.3cm}
	
	In addition, it was shown invertible transformations of dynamical variables can change the number of dynamical degrees of freedom \cite{Jirousek:2022jhh}. In this respect, it is attractive to compare the result with the present work and find a relationship (if any). Besides, in the context of the Lorentz-violating $b^\m$ model, a nonlocal operator was found such that the redefined field satisfies an equation of motion with a scaled coefficient $(1+\xi)b^\m$ \cite{Lehnert:2006id}. It would be nice to investigate whether this result can be achieved by the approach in this work. Finally, since our nonlocal field redefinitions may be understood as a kind of higher-derivative field redefinitions, thus, it is appealing to study our method for theories of higher-derivative terms and also for the nonlocal field theories (see, e.g., \cite{Erbin:2021hkf, Buchbinder:2021rmy} and references therein). 
	

	\section*{Acknowledgments}
	We are grateful to Sylvester James Gates Jr. for useful correspondence and suggestions during the work. The author acknowledges Abasalt Rostami for fruitful discussions. We are also thankful to Mohammad Reza Garousi, Hermann Nicolai, and Shahin Sheikh-Jabbari for reading the manuscript and providing helpful comments, and to Hamid Reza Afshar for his support and encouragement. We also thank Joseph Buchbinder for comments. This work is partially supported by IPM funds.

	%

\end{document}